\documentclass{article}

\usepackage{arxiv}

\usepackage[utf8]{inputenc} 
\usepackage[T1]{fontenc}    
\usepackage{hyperref}       
\usepackage{url}            
\usepackage{booktabs}       
\usepackage{amsfonts}       
\usepackage{nicefrac}       
\usepackage{microtype}      
\usepackage{lipsum}		
\usepackage[authoryear,square,comma]{natbib}
\usepackage{doi}

\usepackage[T1]{fontenc}
\usepackage[pdftex]{graphicx} 
\usepackage{siunitx}

\usepackage{amsmath,mathtools}
\usepackage{amssymb}
\usepackage{amsfonts}
\usepackage{algorithm}
\usepackage[noend]{algpseudocode}
\usepackage{eqparbox}
\usepackage{accents}

\newtheorem{definition}{Definition}
\newtheorem{theorem}{Theorem}
\newtheorem{lemma}{Lemma}

\newtheorem{remark}{Remark}

\newdimen{\algindent}
\algnewcommand\LeftComment[2]{%
\hspace{#1\algindent}$\triangleright$ \eqparbox{COMMENT}{#2} \hfill %
}

\usepackage{graphicx,amsmath}
\newcommand{\colvec}[2][.8]{%
  \scalebox{#1}{%
    \renewcommand{\arraystretch}{.8}%
    $\begin{bmatrix}#2\end{bmatrix}$%
  }
}

\newcommand{\vect}[1]{\mathbf{#1}}
\DeclareMathSymbol{\shortminus}{\mathbin}{AMSa}{"39}

\title{Duality of Stochastic Observability and Constructability and Links to Fisher Information}

\author{\href{https://orcid.org/0000-0002-3268-1467}{\includegraphics[scale=0.06]{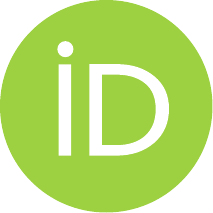}\hspace{1mm}Burak ~Boyacıoğlu} \\
	Department of Mechanical Engineering \\
	University of Nevada, Reno \\
	Reno, NV 89557 \\
	\texttt{bboyacioglu@unr.edu} \\
	\And
	\href{https://orcid.org/0000-0001-6538-7179}{\includegraphics[scale=0.06]{orcid.pdf}\hspace{1mm}Floris ~van Breugel} \\
	Department of Mechanical Engineering \\
	University of Nevada, Reno \\
	Reno, NV 89557 \\
	\texttt{fvanbreugel@unr.edu} \\
}

\renewcommand{\shorttitle}{Duality of Stochastic Observability \& Constructability and Links to Fisher Information}

\hypersetup{
pdftitle={Duality of Stochastic Observability and Constructability and Links to Fisher Information},
pdfauthor={Burak~Boyacioglu, Floris~van~Breugel},
}

\begin{document}
\maketitle

\begin{abstract}
Given a set of measurements, observability characterizes the distinguishability of a system's initial state, whereas constructability focuses on the final state in a trajectory. In the presence of process and/or measurement noise, the Fisher information matrices with respect to the initial and final states—equivalent to the stochastic observability and constructability Gramians—bound the performance of corresponding estimators through the Cramér-Rao inequality. This letter establishes a connection between stochastic observability and constructability of discrete-time linear systems and provides a more numerically robust way for calculating the stochastic observability Gramian. We define a dual system and show that the dual system's stochastic constructability is equivalent to the original system's stochastic observability, and vice versa. This duality enables the interchange of theorems and tools for observability and constructability. For example, we use this result to translate an existing recursive formula for the stochastic constructability Gramian into a formula for recursively calculating the stochastic observability Gramian for both time-varying and time-invariant systems, where this sequence converges for the latter. Finally, we illustrate the robustness of our formula compared to existing (non-recursive) formulas through a numerical example.
\end{abstract}


\section{Introduction}
Over the past several decades, the fields of statistics and control theory have independently developed powerful tools for analyzing dynamic systems, each focusing on extracting reliable information from time-series data. However, the relationship between these conceptually parallel fields has not been comprehensively synthesized in the literature, making it difficult to transfer methods and theoretical insights between fields. This letter bridges this gap by establishing a novel mathematical duality between observability and constructability. We use this duality to derive a robust and efficient recursive formula for calculating stochastic observability Gramians and to make the connections between observability theory, Fisher information, and the posterior Cramér-Rao bound explicitly 
clear.

The Fisher information matrix (FIM), a statistical tool, quantifies the information available for estimating the state of a dynamic system. In real-time applications, state estimates are calculated with an \emph{a posteriori} state estimator, which takes past and present measurements into account. If one has the luxury of waiting for more measurements to accumulate, the estimate can be improved using a smoother \cite{jazwinski1970stochastic}. In both cases, the Cramér-Rao inequality--which involves the FIM--describes how the estimation error is bounded by the system dynamics and outputs, as well as their uncertainties \cite{Manyika1992,crassidis2004}.

Due to the computational burden of calculating the FIM, engineers often analyze \emph{deterministic} observability (ignoring system uncertainties), which typically lead to qualitatively similar conclusions. Observability characterizes the distinguishability of a system's initial state within a trajectory. Quantifying deterministic (un)observability \cite{muller1972,Krener2009,Cellini2023} can help place sensors efficiently and develop active sensing strategies. In the presence of noise, quantification can be based on the \emph{stochastic} observability Gramian, i.e., the FIM with respect to the initial state \cite{Kunwoo2023}. However, existing formulations in the observability literature  for its calculation either ignore process noise \cite{AOKI1968} or (as we illustrate) suffer from numerical instabilities  \cite{tenny2002,LIU2011, SUBASI2014}.

Similar to observability, constructability describes the distinguishability of a system's final state, though it is more rarely discussed as it involves function inverses. The FIM with respect to the final state, aptly called the stochastic constructability Gramian, indicates the performance limit of an \emph{a posteriori} state estimator as it is inversely related to the posterior Cramér-Rao bound \cite{Tichavsky1998}. Avoiding function inverses, Tichavský \emph{et al.} \cite{Tichavsky1998} derive the most general recursive formula for the constructability Gramian as they study discrete-time nonlinear systems with both process and measurement noise.

\begin{figure*}[tb]
\centerline{\includegraphics[width=\linewidth]{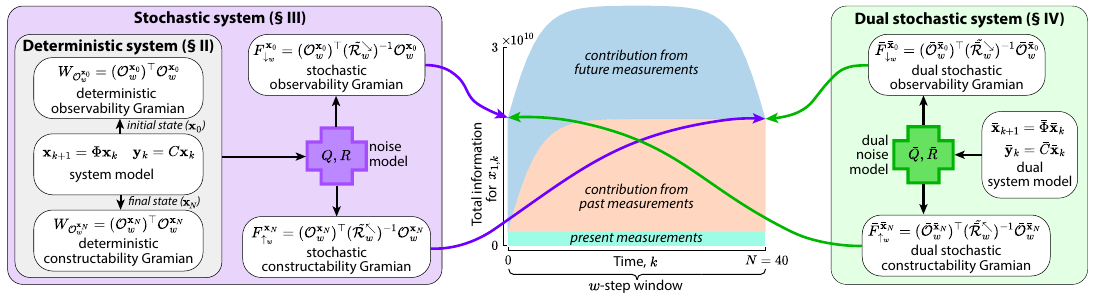}}
\caption{Illustration of key definitions and organization. Middle panel shows the total information for the first state variable of a discrete-time linear time-invariant (LTI) system, i.e., the first diagonal entry of the FIM of all measurements with respect to $\mathbf{x}_k$. The discrete-time LTI system matrices are $\Phi=\colvec[0.5]{ 1&-1\\0&1 }$, $C=\begin{bmatrix}
    1&0
\end{bmatrix}$, $Q=\colvec[0.5]{ 1\times 10^{-11}&-5\times 10^{-18}\\-5\times 10^{-18} &1\times 10^{-17} }$, and $R=2.89\times 10^{-10}$, adapted from \cite[Ex. 3.3]{crassidis2004} to better illustrate the concepts. Present measurement information is constant as $C$ and $R$ are time-invariant, contributing to both observability and constructability: observability reflects the combined information from the future and present, while constructability accounts for past and present measurements. In Sec. \protect\ref{sec:duality}, we construct a dual system in which the stochastic observability Gramian is equivalent to the original system's stochastic constructability Gramian, and vice versa. Due to process noise, distant measurements from the state of interest (whether in the future or past) become irrelevant.
}
\label{fig:outline}
\end{figure*}

In this letter, in search of a recursive formula to obtain the observability Gramian of discrete-time linear time-varying systems with process noise, we present a dual relationship such that the constructability of the dual system is equal to the original system's observability, and vice versa. This duality allows us to translate the recursive stochastic constructability formulas \cite{Tichavsky1998} into the dual system in order to calculate the stochastic observability Gramian of the original system in a numerically more robust way. As one might expect, the dual system is no other than a redefined version of the system propagated backward in time with an invertibility assumption. While such a relationship aligns with intuition given the definitions of observability and constructability, the duality itself has not, to our knowledge, been explicitly stated in the literature. This statement is significant as it creates a clear link between observability and the posterior Cramér-Rao bound, enabling results from one domain to be systematically applied to the other and having the potential to support broader applications, including the interchange of model-free methods for nonlinear systems.

See Fig. \ref{fig:outline} for the organization of the letter.

\section{Background}
There is inconsistency in the literature regarding relevant definitions. For instance, constructability is variously referred to as reconstructability, or with different spellings constructibility and reconstructibility, and sometimes even as on-line observability \cite{AOKI1968}, observability of the final state \cite{sontag1990}, or a future state \cite{Bai2018}, often without acknowledging the ambiguity. In this section, we introduce the terminology and notation used in this letter and relevant background material.

\subsection{Deterministic Observability and Constructability}
The question of observability is whether one can uniquely determine the initial state of a system's trajectory from measurements with the knowledge of inputs, whereas constructability characterizes the distinguishability of the final state given the same measurements and inputs \cite{Hespanha2018}. A typical method to test these system properties is to build the observability and constructability matrices.

Consider the discrete-time linear time-varying (LTV) system dynamics without noise:
\begingroup
\small
 \begin{equation}\label{sysDyn_dtltv}
 \begin{aligned}
     \vect{x}_{k+1} = \Phi_{k+1,k}\vect{x}_k +B_k\vect{u}_k,\quad 
     \vect{y}_k = C_k\vect{x}_k+D_k\vect{u}_k,
 \end{aligned}
\end{equation}\endgroup
where the signals $\vect{x}\in\mathbb{R}^n$, $\vect{u}\in\mathbb{R}^m$, and $\vect{y}\in\mathbb{R}^p$ are the state, input, and output of the system, respectively. $\Phi_{\ell,k}$ denotes the (discrete-time) state transition matrix from $\vect{x}_k$ to $\vect{x}_\ell$, with $\Phi_{k+1,k}$ specifically being the state (or system) matrix at time $k$ and $\Phi_{k,k}$ being the identity matrix. Finally, following \cite{Simon2006}, we assume that the discrete-time system matrix is derived from a matrix exponential with a finite sampling time, hence $\Phi_{k,\ell}^{-1}$ exists.

The initial state, $\vect{x}_0$, can be uniquely obtained from measurements over $w$ time steps, i.e., System (\ref{sysDyn_dtltv}) is $w$-step observable, if and only if the observability matrix,
\begingroup
\small
\begin{equation}\label{observability_w}
    \mathcal{O}_w^{\vect{x}_0}=\begin{bmatrix}C_0\\ C_1\Phi_{1,0}\\\vdots\\ C_{w\shortminus 1}\Phi_{w\shortminus 1,0}\end{bmatrix}\in\mathbb{R}^{wp\times n},
\end{equation}
\endgroup
is full column rank \cite{Bai2018}.

Similarly, the final state in a trajectory, $\vect{x}_N$, is distinguishable in $w$ steps, i.e., the system is $w$-step constructable, if and only if the constructability matrix,
\begingroup
\small
\begin{equation}\label{constructability_w}
    \mathcal{O}^{\vect{x}_N}_w=
    \begin{bmatrix}C_{N}\\ C_{N\shortminus 1}\Phi_{N,N\shortminus 1}^{\shortminus 1}\\\vdots\\ C_{N\shortminus w+1}\Phi_{N,N\shortminus w+1}^{-1} \end{bmatrix}\in\mathbb{R}^{wp\times n},
\end{equation}
\endgroup
is full column rank. In the context of linear systems, observability and constructability do not depend on the input terms. Henceforth, we will omit these terms and focus on autonomous system dynamics,
\begingroup
\small
\begin{equation}\label{sysDyn_adtltv}
  \begin{aligned}
     \vect{x}_{k+1} = \Phi_{k+1,k}\vect{x}_k,\quad 
     \vect{y}_k = C_k\vect{x}_k.
 \end{aligned}
\end{equation}
\endgroup

To understand which directions in the state space are most and least observable, or constructable, it is common to analyze the corresponding Gramians, respectively defined for System (\ref{sysDyn_adtltv}) as:
\begingroup
\small
\begin{align}
    W_{{\mathcal{O}_w^{\vect{x}_0}}} &= \sum_{\tau=0}^{w-1} \Phi_{\tau,0}^\top C_\tau^\top C_\tau \Phi_{\tau,0}=({\mathcal{O}_w^{\vect{x}_0}})^\top\mathcal{O}_w^{\vect{x}_0} \\
    W_{\mathcal{O}^{\vect{x}_N}_w} &= \sum_{\tau=N\shortminus w+1}^{N} \Phi_{N, \tau}^{\shortminus \top} C_\tau^\top C_\tau \Phi_{N,\tau}^{\shortminus 1}=({\mathcal{O}^{\vect{x}_N}_w})^\top \mathcal{O}^{\vect{x}_N}_w,
\end{align}
\endgroup
where $\top$ denotes the matrix transpose.
One can build the initial and final states using the inverse of these $n\times n$ nonnegative-definite matrices, provided they are strictly positive definite \cite{Hespanha2018}. Thus, the condition number and the reciprocal of the minimum eigenvalue of the observability/constructability Gramian can be used as measures of observability or constructability \cite{Krener2009}.

\subsection{Cramér-Rao Bound}
The Cramér-Rao inequality provides a lower bound on the estimation error of \(\vect{x}_k\) given noisy measurements over a $w$-step window, \(\vect{\tilde{Y}}_w\in\mathbb{R}^{wp}\). Estimators that yield error covariances equal to this lower bound are said to be efficient. The inequality is given by
\begingroup
\small
\begin{equation}
P := E\left\{(\hat{\mathbf{x}}_k-\mathbf{x}_k)(\hat{\mathbf{x}}_k-\mathbf{x}_k)^\top\right\} \succeq F^{-1},
\end{equation}
\endgroup
where \(P\) is the error covariance matrix of the estimate \(\hat{\mathbf{x}}_k\), and \(F\) is the Fisher information matrix, defined as:
\begingroup
\small
\begin{equation}
F = E\left\{\left[\frac{\partial}{\partial \mathbf{x}_k} \ln [p({\mathbf{\tilde Y}}_w , \mathbf{x}_k)]\right]\left[\frac{\partial}{\partial \mathbf{x}_k} \ln [p({\mathbf{\tilde Y}}_w , \mathbf{x}_k)]\right]^\top\right\},
\end{equation}
\endgroup
with \(p({\mathbf{\tilde Y}}_w , \mathbf{x}_k)\) being the joint probability density function of \({\mathbf{\tilde Y}}_w\) and \(\mathbf{x}_k\). Thus \(\ln[p({\mathbf{\tilde Y}}_w , \mathbf{x}_k)]\) is the log-likelihood of \({\mathbf{\tilde Y}}_w\) occurring together with \(\mathbf{x}_k\) \cite{crassidis2004}. If $\vect{\tilde Y}_w$ is a linear function of $\vect{x}_k$ with zero-mean Gaussian noise, i.e., if
\begingroup
\small
\begin{equation}
\vect{\tilde Y}_w = H\vect{x}_k+\sum\nolimits_{i}\Upsilon_i\vect{v}_i\in\mathbb{R}^{wp},
\end{equation}
\endgroup
where $\vect{v}_i\sim \mathcal{N}(\vect{0},R_i)$, then the Fisher information matrix can be calculated as:
\begingroup
\small
\begin{equation}\label{eq:linFisher}
    F^{\mathbf{x}_k}_{\vect{\tilde Y}_w}=H^\top {\mathcal{\tilde R}}^{-1}_wH\in\mathbb{R}^{n\times n},
\end{equation}
\endgroup
where ${\mathcal{\tilde R}}_w:=\mathrm{Cov}(\vect{\tilde Y}_w)=\sum_i\Upsilon_i R_i\Upsilon_i^\top$.

In the following section, we will give multiple formulations to calculate the FIM with respect to the initial and final states, i.e., the stochastic observability and constructability Gramians.

\section{Stochastic Observability and Constructability}
This section reviews stochastic observability and constructability Gramians for linear systems, both with and without process noise, and gives their relation to the FIM with respect to the entire state trajectory, i.e., the trajectory information matrix. The Gramian definitions here differ from those derived from empirical approaches \cite{powel2020empirical},
as we neither perturb the state nor run Monte Carlo simulations.

\subsection{Stochastic Systems Without Process Noise}
Consider the autonomous discrete-time LTV system dynamics with measurement noise and without process noise,
\begingroup
\small
\begin{equation}\label{dt-ltv-dynamics-Q0}
  \begin{aligned}
\vect{x}_{k+1} =\Phi_{k+1,k} \vect{x}_k,\quad \vect{\tilde y}_k = C_k\vect{x}_k+\vect{v}_k,
\end{aligned}
\end{equation}\endgroup where $\vect{v}_k\sim\mathcal{N}(\vect{0},R_k)$ and $R_k\succ0$ for all integers $k\geq0$. One can write any noisy measurement $\vect{\tilde{y}}_k$ as a function of the initial state, that is,
\begingroup
\small
\begin{equation}\label{measFunc1}
    \vect{\tilde{y}}_k= C_k\Phi_{k,0}\vect{x}_0+\vect{v}_k.
\end{equation}
\endgroup
Then the collection of the first $w$ measurements is
\begingroup
\small
\begin{equation}
    \vect{\tilde Y}_w^\downarrow:=\begin{bmatrix}
        \vect{\tilde y}_0\\\vdots\\\vect{\tilde y}_{w\shortminus 1}
    \end{bmatrix}=\mathcal{O}_w^{\vect{x}_0}\vect{x}_0+\begin{bmatrix}\vect{v}_0\\\vdots\\\vect{v}_{w\shortminus 1}\end{bmatrix},
\end{equation}
\endgroup
where the arrow denotes the time direction of measurements. Thus, the Fisher information of $\vect{\tilde Y}_w^\downarrow$ with respect to the initial state is
\begingroup
\small
\begin{equation}\label{fisherObs1}
   F_{\downarrow_w}^{\vect{x}_0}=(\mathcal{O}_w^{\vect{x}_0})^\top (\mathcal{R}_w^\searrow)^{-1} \mathcal{O}_w^{\vect{x}_0},
\end{equation}\endgroup
where
\begingroup
\small\begin{equation}
    \mathcal{R}_w^\searrow:=\mathrm{Cov}(\vect{\tilde Y}_w^\downarrow)=\mathrm{blkdiag}({R_0,\dots,R_{w\shortminus 1}}),
\end{equation}\endgroup
and the arrow direction denotes increasing time in the matrix blocks. We will refer to this Fisher information matrix as the $w$\emph{-step stochastic observability Gramian} to make its connection to the traditional definitions of deterministic observability clear. 
In \cite{Kunwoo2023}, where $F_{\downarrow_w}^{\vect{x}_0}$ is called the estimability Gramian, it is noted that $F_{\downarrow_w}^{\vect{x}_0}$ is equal to the deterministic observability Gramian, $W_{{\mathcal{O}_w^{\vect{x}_0}}}$, if the system outputs are scaled such that $R_0=\dots=R_{w\shortminus 1}=I$ where $I$ is the identity matrix.

Although $\mathcal{R}_w^\searrow$ in (\ref{fisherObs1}) is a block diagonal matrix, standard matrix inversion commands do not take advantage of this fact, making calculation of its inverse challenging for large $pw$. Assuming that noise components from different time steps are uncorrelated, i.e., $E\left\{\vect{v}_k\vect{v}_j^\top\right\}=0$ for $k\neq j$, we can obtain $F_{\downarrow_w}^{\vect{x}_0}$ from the recursive formula
\begingroup
\small\begin{equation}
    F_{\downarrow_{k+2}}^{\vect{x}_0}=F_{\downarrow_{k+1}}^{\vect{x}_0}+(C_{k+1}\Phi_{k+1,0})^\top R_{k+1}^{-1}C_{k+1}\Phi_{k+1,0},
\end{equation}\endgroup
initialized with $F_{\downarrow_1}^{\vect{x}_0}=C_0^\top R_0^{-1}C_0$. That is, each measurement makes an independent addition to the information.

Whereas (\ref{measFunc1}) defined each measurement as a function of the initial state, we can alternatively write each measurement as a function of the final state:
\begingroup
\small
\begin{equation}
    \vect{\tilde y}_k=C_k\Phi_{N,k}^{-1}\vect{x}_N+\vect{v}_k,
\end{equation}\endgroup
where $k\leq N$. Then the collection of $w$ measurements, starting with $\vect{\tilde y}_N$, in the reverse-time order would be
\begingroup
\small
\begin{equation}
    \vect{\tilde Y}_w^\uparrow:=\begin{bmatrix}
        \vect{\tilde y}_N\\\vdots\\\vect{\tilde y}_{N\shortminus w+1}
    \end{bmatrix}=\mathcal{O}_w^{\vect{x}_N}\vect{x}_N+\begin{bmatrix}\vect{v}_N\\\vdots\\\vect{v}_{N\shortminus w+1}\end{bmatrix},
\end{equation}\endgroup
and its covariance can be calculated as:
\begingroup
\small
\begin{equation}
    \mathcal{R}_w^\nwarrow:=\mathrm{Cov}(\vect{\tilde Y}_w^\uparrow)=\mathrm{blkdiag}({R_N,\dots,R_{N\shortminus w+1}}).
\end{equation}\endgroup
The Fisher information of $\vect{\tilde Y}_w^\uparrow$ with respect to $\vect{x}_N$ is
\begingroup
\small
\begin{equation}\label{fisherCons1}
   F_{\uparrow_w}^{\vect{x}_N}=(\mathcal{O}_w^{\vect{x}_N})^\top (\mathcal{R}_w^\nwarrow)^{-1} \mathcal{O}_w^{\vect{x}_N},
\end{equation}\endgroup
which we will call as the $w$\emph{-step stochastic constructability Gramian} and note again that it is no different than the deterministic one when output noise levels are normalized, i.e., if $R_N=\dots=R_{N\shortminus w+1}=I$.

Finally, given $w=N+1$, and assuming $E\left\{\vect{v}_k\vect{v}_j^\top\right\}=0$ for $k\neq j$, one can obtain $F_{\uparrow_w}^{\vect{x}_N}$ from the recursive formula
\begingroup
\small
\begin{equation}
    F_{\uparrow_{k+2}}^{\vect{x}_{k+1}}=\Phi_{k+1,k}^{-\top}F_{\uparrow_{k+1}}^{\vect{x}_{k}}\Phi_{k+1,k}^{-1}+C_{k+1}^\top R_{k+1}^{-1}C_{k+1},
\end{equation}\endgroup
initialized with $F_{\uparrow_1}^{\vect{x}_0}=C_0^\top R_0^{-1}C_0$ \cite{ristic2004beyond}.

\subsection{Stochastic Systems with Process Noise}
Unlike the deterministic systems discussed above, in stochastic systems with process noise, the measurements from different time points are correlated due to the propagation of system uncertainty through the system dynamics. This interdependence complicates calculating the stochastic Gramians. Here we provide formulations that will later be replaced by recursive ones, leveraging the duality in terms of the Fisher information.

Consider the autonomous discrete-time LTV system dynamics with process and measurement noise,
\begingroup
\small
\begin{equation}\label{dt-ltv-dynamics}
  \begin{aligned}
\vect{x}_{k+1} =\Phi_{k+1,k} \vect{x}_k+\vect{w}_k,\quad \vect{\tilde y}_k = C_k\vect{x}_k+\vect{v}_k,
\end{aligned}
\end{equation}\endgroup
where $\vect{w}_k\sim\mathcal{N}(\vect{0},Q_k)$, $\vect{v}_k\sim\mathcal{N}(\vect{0},R_k)$, and $Q_k, R_k\succ0$ for all integers $k$.
Given the vector of the first $w$ measurements for System (\ref{dt-ltv-dynamics}), $\vect{\tilde Y}_w^\downarrow$, the $w$-step stochastic observability Gramian is
\begingroup
\small
\begin{equation}\label{fisherObs2}
   F_{\downarrow_w}^{\vect{x}_0}=(\mathcal{O}_w^{\vect{x}_0})^\top ({\mathcal{\tilde R}}^\searrow_w)^{-1} \mathcal{O}_w^{\vect{x}_0},
\end{equation}\endgroup
where
\begingroup
\small
\begin{equation}\label{fisherObs2b}
  {\mathcal{\tilde R}}^\searrow_w:=\mathrm{Cov}(\vect{\tilde Y}^\downarrow_w)=\begin{bmatrix}
        R_0&0_{p\times p} &\cdots&0_{p\times p}\\
        0_{p\times p}& R_{2,2}& \cdots&R_{2,w}\\
        \vdots&\vdots&\ddots&\vdots\\
        0_{p\times p}& R_{w,2}& \cdots&R_{w,w}
    \end{bmatrix},
\end{equation}
\endgroup
with lower triangle block matrices,
\begingroup
\scriptsize
\begin{equation}\label{fisherObs2c}
    \stepcounter{equation} 
    R_{j+1,k+1}= \begin{cases} 
      R_j+\sum_{i=1}^{k} C_j\Phi_{j,i}Q_{i\shortminus 1}(C_j\Phi_{j,i})^\top &  \text{if }j=k>0 \\
      \sum_{i=1}^k C_j\Phi_{j,i}Q_{i\shortminus 1}(C_k\Phi_{k,i})^\top &  \text{if }j>k>0,
   \end{cases}
   \tag*{\normalsize(\small\theequation)}
\end{equation}
\endgroup
and the rest determined from symmetry
, which can be seen by expressing the measurements \( \mathbf{y}_j \) and \( \mathbf{y}_k \) in terms of \( \mathbf{x}_0 \), with \( R_{j+1,k+1} \) representing the covariance matrix of these two output vectors. Equation (\ref{fisherObs2}) then follows from (\ref{eq:linFisher}).

Here, the formulation of ${\mathcal{\tilde R}}^\searrow_w$, although correct, is not computationally efficient to implement because of the sums required in each block. An alternative method, given in \cite{tenny2002,LIU2011, SUBASI2014}, first expresses $\vect{\tilde Y}^\downarrow_w$ as a linear function of $\vect{x}_0$ and the noise components, then uses the linear propagation of uncertainty. We provide this arguably more intuitive formula here, but refer the reader to \cite{LIU2011} for the explicit form. The covariance matrix of the set of measurements is
\begingroup
\small
\begin{equation}
 {\mathcal{\tilde R}}^\searrow_w={\mathcal{R}}_w^\searrow+M_w^{\vect{x}_0} \mathcal{Q}_w^\searrow (M_w^{\vect{x}_0})^\top,
\end{equation}\endgroup
where $\mathcal{Q}_w^\searrow=\mathrm{blkdiag}(Q_0,\dots,Q_{w\shortminus 2},*_{n\times n})$, and the $wp\times wn$ matrix $M_w^{\vect{x}_0}$ for $i, j \leq w$ is built using the block matrices
\begingroup
\small
\begin{equation*}
[M_w^{\vect{x}_0}]_{i,j}= \begin{cases}0_{p\times n} & \text { if } j \geq i \\ 
C_{i-1}\Phi_{i-1,j}   & \text { otherwise. }\end{cases}
\end{equation*}\endgroup
Here, \( *_{n \times n} \) denotes any \( n \times n \) matrix. Note that \( M_w^{\vect{x}_0} \) could be \( wp \times (w-1)n \), allowing \( \mathcal{Q}_w^\searrow \) to omit this last block, but we chose consistency with \cite{tenny2002}.

Following similar steps, one can build the $w$-step stochastic constructability Gramian. Let $\vect{\tilde Y}_w^\uparrow$ be the vector of the last $w$ measurements in the reverse-time order for System (\ref{dt-ltv-dynamics}). Then the FIM with respect to the final state is given by
\begingroup
\small 
\begin{equation}\label{fisherCon2}
   F_{\uparrow_w}^{\vect{x}_N}=(\mathcal{O}_w^{\vect{x}_N})^\top (\mathcal{\tilde R}^\nwarrow_w)^{-1} \mathcal{O}_w^{\vect{x}_N},
\end{equation}\endgroup
where
\begingroup
\small 
\begin{gather}
  \mathcal{\tilde R}^\nwarrow_w:=\mathrm{Cov}(\vect{\tilde Y}^\uparrow_w)=\mathcal{R}_w^\nwarrow+M_w^{\vect{x}_N} \mathcal{Q}_w^\nwarrow (M_w^{\vect{x}_N})^\top,\\
  \mathcal{R}_w^\nwarrow=\mathrm{blkdiag}(R_{N},\dots,R_{N\shortminus w+1}),\notag\\
  \mathcal{Q}_w^\nwarrow=\mathrm{blkdiag}(Q_{N\shortminus 1},\dots,Q_{N\shortminus w+1},*_{n\times n}),\notag
\end{gather}\endgroup 
and $M_w^{\vect{x}_N}$ for $i, j \leq w$ is built using the block matrices
\begingroup
\small 
\begin{equation*}
[M_w^{\vect{x}_N}]_{i, j}= \begin{cases}0_{p\times n} & \text { if } j \geq i \\ 
C_{N\shortminus i+1}\Phi_{N\shortminus j+1,N\shortminus i+1}^{-1} & \text { otherwise. }\end{cases}
\end{equation*}\endgroup

One can calculate the output covariance matrix $\mathcal{\tilde R}^\nwarrow_w$ similar to (\ref{fisherObs2}-25). Because $\mathcal{\tilde R}^\searrow_w$ and $\mathcal{\tilde R}^\nwarrow_w$ are large non-block-diagonal $wp$-by-$wp$ matrices, inverting them can introduce significant numerical errors and require substantial memory and computation time.



\subsection{Relation to Trajectory Information Matrix}
The trajectory information matrix is the Fisher information matrix of a set of noisy measurements with respect to the state trajectory of a dynamic system, and its inverse sets a lower bound for the estimation error of any state in the trajectory when the entire set of measurements is available. In this subsection, we show the trajectory information matrix's relation to stochastic observability and constructability Gramians and explore the interpretation of this connection.

Let $\vect{X}_w$ be the collection of state vectors from $k=0$ to $k=w-1$. Then the Fisher information matrix of $\vect{\tilde Y}_w$ with respect to $\vect{X}_w$ is a $wn$-by-$wn$ matrix such that
\begingroup
\small
\begin{equation}
    \mathcal{F}_w^{-1}=\begin{bmatrix}
        (F_{\downarrow_w}^{\vect{x}_0})^{-1}&\mathcal{P}_{1,2}&\cdots&\mathcal{P}_{1,w}\\
        \mathcal{P}_{2,1}&\mathcal{P}_{2,2}
        &\cdots&\mathcal{P}_{2,w}\\
\vdots&\vdots&\ddots&\vdots\\  \mathcal{P}_{w,1}&\mathcal{P}_{w,2}&\cdots&(F_{\uparrow_w}^{\vect{x}_{w\shortminus 1}})^{-1}
    \end{bmatrix},
\end{equation}
\endgroup
where $\mathcal{P}_{i,j}$'s are the error covariance matrix blocks of the trajectory. Notice that the stochastic observability and constructability Gramians appear in the upper left and lower right corners of $\mathcal{F}_w^{-1}$, respectively.  From the structure of $\mathcal{F}_w$ and $\mathcal{F}_{w+1}$, a recursive formula to calculate $F_{\uparrow_{N+1}}^{\vect{x}_{N}}$ is given as:
\begingroup
\small 
\begin{equation}\label{eq_rec3}
\begin{aligned}
    F_{\uparrow_{k+2}}^{\vect{x}_{k+1}} =&\shortminus Q_k^{\shortminus 1}\Phi_{k+1,k} (F_{\uparrow_{k+1}}^{\vect{x}_{k}}+\Phi_{k+1,k}^\top Q_k^{\shortminus 1}\Phi_{k+1,k})^{\shortminus 1}\Phi_{k+1,k}^\top Q_k^{\shortminus 1}+Q_k^{\shortminus 1}+C_{k+1}^\top R_{k+1}^{\shortminus 1}C_{k+1},
\end{aligned}
\end{equation}\endgroup 
where $F_{\uparrow_{1}}^{\vect{x}_{0}}=C_0R_0^{-1}C_0$ \cite{crassidis2004}. That is, we now have $n$-by-$n$ matrix inverses, avoiding the inversion of the larger $wp$-by-$wp$ matrix in (\ref{fisherCon2}). However, there is no recursive formula to obtain $F_{\downarrow_w}^{\vect{x}_0}$. Finally, we note that the inverse of the other diagonal elements of $\mathcal{F}_w^{-1}$—the Fisher information of $\vect{\tilde Y}_w$ with respect to intermediate states—can be derived from the sum of their observability and constructability Gramians, minus the information at time $k$, since both Gramians incorporate that information.

\section{Duality of Stochastic Observability and Constructability}\label{sec:duality}

In this section, we introduce a dual discrete-time linear system to System (\ref{dt-ltv-dynamics}) in terms of stochastic observability and constructability, where $\vect{\bar \bullet}$ denotes a dual system variable to $\vect{\bullet}$. Using this dual system we present a recursive formula to calculate the stochastic observability Gramian, which is missing from the literature. We then give more conclusive results, including convergence, for time-invariant systems.

\subsection{Linear Time-varying Systems}
We start with defining duality in the context of observability and constructability for time-varying systems.

\begin{definition}\label{defDual}
    Given $w=N+1$, two stochastic, time-varying systems are $w$\emph{-step dual in terms of observability and constructability} if the $w$-step stochastic constructability Gramian of the first system is equal to the $w$-step observability Gramian of the second system, and vice versa.
\end{definition}

Consider the autonomous discrete-time LTV system dynamics with process and measurement noise,
\begingroup
\small 
\begin{equation}\label{dt-dual-ltv-dynamics}
  \begin{aligned}
\vect{\bar x}_{k+1} =\bar\Phi_{k+1,k} \vect{\bar x}_k+\vect{\bar w}_k,\quad \vect{\tilde {\bar y}}_k = \bar C_k\vect{\bar x}_k+\vect{\bar v}_k,
\end{aligned}
\end{equation}\endgroup
where $\vect{\bar w}_k\sim\mathcal{N}(\vect{0},\bar Q_k)$, $\vect{\bar v}_k\sim\mathcal{N}(\vect{0},\bar R_k)$, and $\bar Q_k, \bar R_k\succ0$ for all integers $k$.
\begin{theorem}\label{theo_ltv}
    Let the dynamics of Systems (\ref{dt-ltv-dynamics}) and (\ref{dt-dual-ltv-dynamics}) be related such that $\bar \Phi_{N,N\shortminus1}=\Phi^{-1}_{1,0}$, $\bar \Phi_{N\shortminus1,N\shortminus2}=\Phi^{-1}_{2,1}$, \dots, $\bar C_{N}=C_0$, $\bar C_{N\shortminus1}=C_1$, \dots, $\bar Q_{N\shortminus1}=\Phi^{-1}_{1,0}Q_0\Phi^{-\top}_{1,0}$, $\bar Q_{N\shortminus2}=\Phi^{-1}_{2,1}Q_1\Phi^{-\top}_{2,1}$, \dots, and $\bar R_{N}=R_0$, $\bar R_{N\shortminus1}=R_1$, \dots. Then Systems (\ref{dt-ltv-dynamics}) and (\ref{dt-dual-ltv-dynamics}) are dual in terms of observability and constructability.
\end{theorem}
\emph{Proof.} Given $N=w-1$, we first show that $w$-step constructability matrix of System (\ref{dt-dual-ltv-dynamics}), $\mathcal{\bar O}_w^{\vect{\bar x}_N}$, is equal to $\mathcal{O}_w^{\vect{x}_0}$:
\begingroup
\small
\begin{equation*}
    \mathcal{\bar O}_w^{\vect{\bar x}_N}=\begin{bmatrix}\bar C_{N}\\ \bar C_{N\shortminus 1}\bar \Phi_{N,N\shortminus 1}^{-1}\\\vdots\\ \bar C_{0}\bar \Phi_{N,0}^{-1} \end{bmatrix}=\begin{bmatrix}C_0\\ C_1\Phi_{1,0}\\\vdots\\ C_{w\shortminus 1}\Phi_{w\shortminus 1,0}\end{bmatrix}=\mathcal{O}_w^{\vect{x}_0}.
\end{equation*}\endgroup
Second, we show that the measurement covariance matrices are the same. Let
\begingroup
\small
\begin{equation}
  \mathcal{\tilde {\bar R}}^\nwarrow_w:=\mathrm{Cov}(\vect{\tilde {\bar Y}}^\uparrow_w)=\mathcal{\bar R}_w^\nwarrow+\bar M_w^{\vect{\bar x}_N} \mathcal{\bar Q}_w^\nwarrow (\bar M_w^{\vect{\bar x}_N})^\top,\end{equation}\endgroup
where $\mathcal{\bar R}_w^\nwarrow=\mathrm{blkdiag}({\bar R_N,\dots,\bar R_{0}})$, $\mathcal{\bar Q}_w^\nwarrow=\mathrm{blkdiag}(\bar Q_{N\shortminus 1},\dots,\bar Q_{0},*_{n\times n})$, and the matrix $\bar M_w^{\vect{\bar x}_N}$ for $i, j \leq w$ is built using the block matrices
\begingroup
\small
\begin{equation*}
[\bar M_w^{\vect{\bar x}_N}]_{i, j}= \begin{cases}0_{p\times n} & \text { if } j \geq i \\ 
\bar C_{N\shortminus i+1}\bar \Phi_{N\shortminus j+1,N\shortminus i+1}^{-1} & \text { otherwise. }\end{cases}
\end{equation*}\endgroup
It can be readily observed that $\mathcal{\bar R}_w^\nwarrow=\mathcal{R}_w^\searrow$, and although $\mathcal{\bar Q}_w^\nwarrow\neq \mathcal{Q}_w^\searrow$, the resulting matrix products would give
\begingroup
\small
\begin{equation*}
    \bar M_w^{\vect{\bar x}_N} \mathcal{\bar Q}_w^\nwarrow  (\bar M_w^{\vect{\bar x}_N})^\top=M_w^{\vect{x}_0} \mathcal{Q}_w^\searrow (M_w^{\vect{x}_0})^\top.
\end{equation*}\endgroup
Therefore $\mathcal{\tilde {\bar R}}^\nwarrow_w=\mathcal{\tilde {R}}^\searrow_w$ and $\bar F_{\uparrow_w}^{\vect{\bar x}_N}=F_{\downarrow_w}^{\vect{x}_0}$. Similarly, one can show that $\bar F_{\downarrow_w}^{\vect{\bar x}_0}=F_{\uparrow_w}^{\vect{x}_N}$.\hfill $\blacksquare$

\begin{remark}\label{rem1}
    Definition \ref{defDual} implies that the number of time steps considered for observability matters when deriving the dual system. Indeed, the constructability of the dual system evolves differently than the observability of the original system, although the final matrices are the same. 
     On the other hand, one can show that $\bar F_{\uparrow_{w\shortminus 1}}^{\vect{\bar x}_{N\shortminus 1}}= F_{\downarrow_{w\shortminus 1}}^{\vect{x}_1}$, $\bar F_{\uparrow_{w\shortminus 2}}^{\vect{\bar x}_{N\shortminus 2}}= F_{\downarrow_{w\shortminus 2}}^{\vect{x}_2}$, \dots, $\bar F_{\uparrow_{1}}^{\vect{\bar x}_0}= F_{\downarrow_{1}}^{\vect{x}_N}$ .
\end{remark}

Now that we have established the duality of the two systems, in light of Remark \ref{rem1}, it is possible to obtain $ F_{\downarrow_w}^{\vect{ x}_0}$ recursively.
\begin{lemma}\label{lem1}
    Given System (\ref{dt-ltv-dynamics}), the stochastic observability Gramian, $F_{\downarrow_{w}}^{\vect{x}_0}$, can be obtained using
\begingroup
\small
\begin{equation}\label{eq_rec4}
\begin{aligned}F_{\downarrow_{k+2}}^{\vect{x}_{N\shortminus k\shortminus 1}} =&\shortminus \phi^\top Q_{N\shortminus k\shortminus 1}^{\shortminus 1}
    (F_{\downarrow_{k+1}}^{\vect{x}_{N\shortminus k}}+Q_{N\shortminus k\shortminus 1}^{\shortminus 1})^{\shortminus 1}Q_{N\shortminus k\shortminus 1}^{\shortminus 1}\phi+\phi^\top Q_{N\shortminus k\shortminus 1}^{\shortminus 1}\phi+C_{N\shortminus k\shortminus 1}^\top R_{N\shortminus k\shortminus 1}^{\shortminus 1}C_{N\shortminus k\shortminus 1},
\end{aligned}
\end{equation}\endgroup
with $F_{\downarrow_1}^{\vect{x}_N}=C_NR_N^{-1}C_N$ where $\phi=\Phi^\top_{N\shortminus k,N\shortminus k\shortminus 1}$.
\end{lemma}
\emph{Proof.} We first write (\ref{eq_rec3}) for the dual system (\ref{dt-dual-ltv-dynamics}) as:
\begingroup
\small
\begin{equation*}
\begin{aligned}
    \bar F_{\uparrow_{k+2}}^{\vect{\bar x}_{k+1}} =&\shortminus \bar Q_k^{\shortminus 1}\bar \phi (\bar F_{\uparrow_{k+1}}^{\vect{\bar x}_{k}}+\bar \phi^\top \bar Q_k^{\shortminus 1}\bar \phi)^{\shortminus 1}\bar \phi^\top \bar Q_k^{\shortminus 1}+\bar Q_k^{\shortminus 1}+\bar C_{k+1}^\top \bar R_{k+1}^{-1}\bar C_{k+1},
\end{aligned}
\end{equation*}\endgroup
where $\bar F_{\uparrow_{1}}^{\vect{\bar x}_{0}}=\bar C_0\bar R_0^{-1}\bar C_0$ and $\bar \phi=\bar \Phi_{k+1,k}$,
then replace the dual system matrices applying their given relationship to the original system matrices.
\hfill $\blacksquare$

\begin{remark}
    The $wp$-by-$wp$ matrix inversion in (\ref{fisherObs2}) is avoided by Lemma \ref{lem1}. Additionally, the system matrix inverse, though used in the derivation, does not appear in (\ref{eq_rec4}).
\end{remark}

\subsection{Linear Time-invariant Systems}
Now we discuss the implications of duality for the special case of time-invariant systems.

\begin{definition}\label{defDualLTI}
    For any number of steps $w$, two stochastic, time-invariant systems are \emph{dual in terms of observability and constructability} if the $w$-step stochastic constructability Gramian of the first system is equal to the $w$-step observability Gramian of the second system, and vice versa.
\end{definition}

Consider the dynamics of two autonomous discrete-time LTI systems with process and measurement noise,
\begingroup
\small
\begin{equation}\label{dt-lti-dynamics}
  \begin{aligned}
\vect{x}_{k+1} =\Phi \vect{x}_k+\vect{w}_k,\quad \vect{\tilde {y}}_k = C\vect{x}_k+\vect{v}_k,
\end{aligned}
\end{equation}\endgroup
and
\begingroup
\small
\begin{equation}\label{dt-dual-lti-dynamics}
  \begin{aligned}
\vect{\bar x}_{k+1} =\bar\Phi \vect{\bar x}_k+\vect{\bar w}_k,\quad \vect{\tilde {\bar y}}_k = \bar C\vect{\bar x}_k+\vect{\bar v}_k,
\end{aligned}
\end{equation}\endgroup
where $\vect{w}_k\sim\mathcal{N}(\vect{0},Q)$, $\vect{v}_k\sim\mathcal{N}(\vect{0},R)$, $\vect{\bar w}_k\sim\mathcal{N}(\vect{0},\bar Q)$, $\vect{\bar v}_k\sim\mathcal{N}(\vect{0},\bar R)$, and $Q,R,\bar Q, \bar R\succ0$ for all integers $k$. 

\begin{theorem}
    Let the dynamics of Systems (\ref{dt-lti-dynamics}) and (\ref{dt-dual-lti-dynamics}) be related such that $\bar \Phi=\Phi^{-1}$, $\bar C=C$, $\bar Q=\Phi^{-1}Q\Phi^{-\top}$, and $\bar R=R$. Then Systems (\ref{dt-lti-dynamics}) and (\ref{dt-dual-lti-dynamics}) are dual in terms of observability and constructability.
\end{theorem}

\emph{Proof.} This result follows directly from Theorem \ref{theo_ltv} and the duality definition for linear time-invariant systems.\hfill $\blacksquare$

Now that we have illustrated the duality of the two systems, we can write a recursive formula, similar to (\ref{eq_rec4}), for time-invariant dynamics:
\begingroup
\small
\begin{equation}\label{eq_rec4.5}
\begin{aligned}
    F_{\downarrow_{k+2}}^{\vect{x}_{N\shortminus k\shortminus 1}} =&\shortminus \Phi^\top Q^{\shortminus 1}
    (F_{\downarrow_{k+1}}^{\vect{x}_{N\shortminus k}}+Q^{\shortminus 1})^{\shortminus 1}Q^{\shortminus 1}\Phi+\Phi^\top Q^{\shortminus 1}\Phi+ C^\top  R^{\shortminus 1} C,
\end{aligned}
\end{equation}\endgroup
with $F_{\downarrow_1}^{\vect{x}_N}=CR^{-1}C$. Time-invariance also implies
\begingroup
\small
\begin{equation}\label{eq_rec5}
\begin{aligned}
    F_{\downarrow_{k+2}}^{\vect{x}_{0}} =&\shortminus \Phi^\top Q^{\shortminus 1}
    (F_{\downarrow_{k+1}}^{\vect{x}_{0}}+Q^{\shortminus 1})^{\shortminus 1}Q^{\shortminus 1}\Phi+\Phi^\top Q^{\shortminus 1}\Phi+ C^\top  R^{\shortminus 1} C,
\end{aligned}
\end{equation}\endgroup
with $F_{\downarrow_1}^{\vect{x}_0}=CR^{-1}C$ since LTI observability does not depend on the system state. Finally, we observe that the stochastic observability Gramian converges to a matrix $F_{\downarrow_{\infty}}^{\vect{x}_{0}}$, which can be expressed from the Riccati equation
\begingroup
\small
\begin{equation*}
F_{\downarrow_{\infty}}^{\vect{x}_{0}} = \shortminus \Phi^\top Q^{\shortminus 1}(F_{\downarrow_{\infty}}^{\vect{x}_{0}} +Q^{\shortminus 1})^{\shortminus 1}Q^{\shortminus 1}\Phi
+\Phi^\top Q^{\shortminus 1}\Phi+C^\top R^{\shortminus 1}C,
\end{equation*}\endgroup
or equivalently,
\begingroup
\small
\begin{equation}\label{eq:ric}
\begin{aligned}
F_{\downarrow_{\infty}}^{\vect{x}_{0}} = &\Phi^\top F_{\downarrow_{\infty}}^{\vect{x}_{0}}  \Phi\shortminus  \Phi^\top F_{\downarrow_{\infty}}^{\vect{x}_{0}}  (F_{\downarrow_{\infty}}^{\vect{x}_{0}} +Q^{\shortminus 1})^{\shortminus 1} ( \Phi^\top F_{\downarrow_{\infty}}^{\vect{x}_{0}} )^\top+C^\top R^{\shortminus 1}C.
\end{aligned}
\end{equation}
\endgroup
This second equation can be solved using MATLAB's \texttt{idare} function  or an equivalent command. The convergence implies that, unlike in system models without process noise, as measurements become more distant from $\vect{x}_0$, their contribution to the information becomes insignificant.

\section{Numerical Example}
In this section, we illustrate how the recursive formula (\ref{eq_rec4}) obtained from duality to calculate the stochastic observability Gramian outperforms both the existing formula and the formula presented in (\ref{fisherObs2}-25) in terms of numerical robustness and memory requirements. We study a discrete-time LTV system with the dynamics
\begingroup
\small
\begin{equation*}
  \begin{aligned}
\vect{x}_{k+1} &=\begin{bmatrix}
    2 &-1+\sin(k\pi/18)\\\cos(k\pi/18) &1
\end{bmatrix} \vect{x}_k+\vect{w}_k\\ {\tilde y}_k &=\begin{bmatrix}
    1&0
\end{bmatrix}\vect{x}_k+v_k,
\end{aligned}
\end{equation*}\endgroup
where $\mathbf{w}_k\sim \mathcal{N}(\mathbf{0},\colvec[0.5]{ 3.6\times10^{-2}&1.2\times10^{-2}\\1.2\times10^{-2} &6\times10^{-2} })$ and $v_k\sim \mathcal{N}(0,0.1)$. We obtain the stochastic observability Gramian for an increasing window size from 1 to 31.

\begin{figure}[tbh]
\centerline{\includegraphics[width=0.5\linewidth]{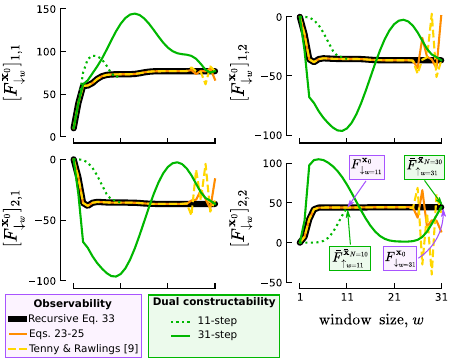}}
\caption{Comparison of three approaches for calculating the stochastic observability Gramian highlighting the numerical stability of our recursive formulation. Each panel shows one of the four entries of the stochastic observability Gramian over time, sharing the same x-axis label. The 11- and 31-step dual system's constructability Gramians entries are also shown to illustrate that they reach the same final value as $F^{\mathbf{x}_0}_{\downarrow_w}$, despite differences in intermediate values. Although the observability calculations for each $w$  require unique dual systems, these calculations can be run in parallel. Negative values of the off-diagonal entries of $F^{\mathbf{x}_0}_{\downarrow_w}$ are acceptable as long as $F^{\mathbf{x}_0}_{\downarrow_w}$ remains positive definite.}
\label{fig:sims}
\end{figure}

Figure \ref{fig:sims} shows the computation results. Numerical instabilities in the non-recursive formulas become apparent after $w=25$ as $[F^{\mathbf{x}_0}_{\downarrow_w}]_{1,2}$ and $[F^{\mathbf{x}_0}_{\downarrow_w}]_{2,1}$ diverge, breaking the symmetry of a FIM, while $[F^{\mathbf{x}_0}_{\downarrow_w}]_{1,1}$ and $[F^{\mathbf{x}_0}_{\downarrow_w}]_{2,2}$ get values that contradict their non-decreasing nature.  Although we later avoided the explicit inverse in (\ref{fisherObs2}) by solving the system $\mathcal{\tilde R}^\searrow_w X=\mathcal{O}_w^{\vect{x}_0}$, numerical instabilities persisted. Those results, along with the code for all three methods, are available on our GitHub \cite{paperGitHub}. Lastly, for a large window size like \(w = 1000\), \(\mathcal{\tilde R}^\searrow_w\)  of a system with one output ($p=1$) requires 80 MB of storage with non-recursive formulas, assuming double precision (8 B/scalar), whereas (\ref{eq_rec4}) has no significant memory requirements.

\section{Conclusions}
We have established the duality of observability and constructability in the presence of process and measurement noise by building their respective Fisher information matrices. This duality serves as a bridge between the existing literature on stochastic observability and the posterior Cramér-Rao bound, allowing results from each domain to be applied to the other. We used this relationship to derive a recursive formula for stochastic observability from the existing recursive formula for stochastic constructability, as demonstrated in Lemma \ref{lem1}. In retrospect, this formulation does appear in the derivation of the forward-backward smoother without any explicit reference to the observability Gramian \cite{crassidis2004,Simon2006,Giuseppe1992}. 
Our duality result, which is not linear algebraic (unlike the duality of observability and controllability), conceptually supports method interchangeability, such as addressing the singularity of $Q$ in (\ref{eq_rec4}) using \cite{Tichavsky1998} or extending data-driven observability methods, like \cite{Massiani2024}, to posterior Cramér-Rao bound analysis. Ultimately, we hope that this letter will spark further studies on relationships among observability, constructability, and information theory in the context of stochastic systems.
\section*{Acknowledgement}
The authors thank Richard Murray, Nick Andrews, Natalie Brace, and Ben Cellini for feedback on the manuscript.

\section*{Funding Sources}
This work was supported by the National Science Foundation AI Institute in Dynamic Systems (2112085).

\bibliographystyle{plainnat}
\bibliography{biblio}  

\begin{thebibliography}{21}
\providecommand{\natexlab}[1]{#1}
\providecommand{\url}[1]{\texttt{#1}}
\expandafter\ifx\csname urlstyle\endcsname\relax
  \providecommand{\doi}[1]{doi: #1}\else
  \providecommand{\doi}{doi: \begingroup \urlstyle{rm}\Url}\fi

\bibitem[Aoki(1968)]{AOKI1968}
Masanao Aoki.
\newblock On observability of stochastic discrete-time dynamic systems.
\newblock \emph{J. of the Franklin Institute}, 286\penalty0 (1):\penalty0 36--58, 1968.
\newblock ISSN 0016-0032.
\newblock \doi{https://doi.org/10.1016/0016-0032(68)90107-5}.

\bibitem[Bai and Taylor(2018)]{Bai2018}
He~Bai and Clark~N. Taylor.
\newblock Observability driven path planning for relative navigation of unmanned aerial systems.
\newblock In \emph{2018 IEEE/ION Position, Location and Navigation Symp.}, pages 793--800, 2018.
\newblock \doi{10.1109/PLANS.2018.8373455}.

\bibitem[Boyacıoğlu(2024)]{paperGitHub}
Burak Boyacıoğlu.
\newblock Stochastic linear observability and constructability ({SLOC}).
\newblock \url{https://github.com/boyacioglub/SLOC}, 2024.

\bibitem[Cellini et~al.(2023)Cellini, Boyacıoğlu, and Van~Breugel]{Cellini2023}
Benjamin Cellini, Burak Boyacıoğlu, and Floris Van~Breugel.
\newblock Empirical individual state observability.
\newblock In \emph{Proc. of the 62nd IEEE Conf. on Decision and Control}, pages 8450--8456, 2023.
\newblock \doi{10.1109/CDC49753.2023.10383812}.

\bibitem[Crassidis and Junkins(2012)]{crassidis2004}
John~L Crassidis and John~L Junkins.
\newblock \emph{Optimal Estimation of Dynamic Systems}.
\newblock Chapman and Hall/CRC, 2nd edition, 2012.
\newblock \doi{10.1201/9780203509128}.

\bibitem[De~Nicolao(1992)]{Giuseppe1992}
Giuseppe De~Nicolao.
\newblock On the time-varying {R}iccati difference equation of optimal filtering.
\newblock \emph{SIAM J. on Control and Optimization}, 30\penalty0 (6):\penalty0 1251--1269, 1992.
\newblock \doi{10.1137/0330066}.

\bibitem[Hespanha(2018)]{Hespanha2018}
João~P. Hespanha.
\newblock \emph{Linear Systems Theory}.
\newblock Princeton University Press, 2nd edition, 2018.
\newblock ISBN 9780691179575.

\bibitem[Jazwinski(1970)]{jazwinski1970stochastic}
Andrew~H. Jazwinski.
\newblock \emph{Stochastic Processes and Filtering Theory}.
\newblock Academic Press, 1970.
\newblock ISBN 9780124157439.

\bibitem[Krener and Ide(2009)]{Krener2009}
Arthur~J. Krener and Kayo Ide.
\newblock {Measures of unobservability}.
\newblock In \emph{Proc. of the 48h IEEE Conf. on Decision and Control held jointly with 28th Chinese Control Conf.}, pages 6401--6406, 2009.
\newblock \doi{10.1109/CDC.2009.5400067}.

\bibitem[Kunwoo et~al.(2023)Kunwoo, Umezu, Konno, and Kashima]{Kunwoo2023}
Lee Kunwoo, Yusuke Umezu, Kaiki Konno, and Kenji Kashima.
\newblock Observability {Gramian} for {B}ayesian inference in nonlinear systems with its industrial application.
\newblock \emph{IEEE Control Systems Letters}, 7:\penalty0 871--876, 2023.
\newblock \doi{10.1109/LCSYS.2022.3227452}.

\bibitem[Liu and Bitmead(2011)]{LIU2011}
Andrew~R. Liu and Robert~R. Bitmead.
\newblock Stochastic observability in network state estimation and control.
\newblock \emph{Automatica}, 47\penalty0 (1):\penalty0 65--78, 2011.
\newblock \doi{https://doi.org/10.1016/j.automatica.2010.10.017}.

\bibitem[Manyika and Durrant-Whyte(1992)]{Manyika1992}
James~M Manyika and Hugh~F Durrant-Whyte.
\newblock Information-theoretic approach to management in decentralized data fusion.
\newblock In \emph{Sensor Fusion V}, volume 1828, pages 202--213. SPIE, 1992.
\newblock \doi{10.1117/12.131652}.

\bibitem[Massiani et~al.(2024)Massiani, Buisson-Fenet, Solowjow, Meglio, and Trimpe]{Massiani2024}
Pierre-François Massiani, Mona Buisson-Fenet, Friedrich Solowjow, Florent~Di Meglio, and Sebastian Trimpe.
\newblock Data-driven observability analysis for nonlinear stochastic systems.
\newblock \emph{IEEE Trans. on Automatic Control}, 69\penalty0 (6):\penalty0 4042--4049, 2024.
\newblock \doi{10.1109/TAC.2023.3346812}.

\bibitem[Müller and Weber(1972)]{muller1972}
Peter~C. Müller and Hans~I. Weber.
\newblock Analysis and optimization of certain qualities of controllability and observability for linear dynamical systems.
\newblock \emph{Automatica}, 8\penalty0 (3):\penalty0 237--246, 1972.
\newblock \doi{10.1016/0005-1098(72)90044-1}.

\bibitem[Powel and Morgansen(2020)]{powel2020empirical}
Nathan Powel and Kristi~A. Morgansen.
\newblock Empirical observability {Gramian} for stochastic observability of nonlinear systems.
\newblock \url{https://arxiv.org/abs/2006.07451}, 2020.

\bibitem[Ristic et~al.(2004)Ristic, Arulampalam, and Gordon]{ristic2004beyond}
Branko Ristic, Sanjeev Arulampalam, and Neil Gordon.
\newblock \emph{Beyond the Kalman Filter: Particle Filters for Tracking Applications}.
\newblock Artech House, 2004.
\newblock ISBN 9781580536318.

\bibitem[Simon(2006)]{Simon2006}
Dan Simon.
\newblock \emph{Optimal State Estimation: {K}alman, H$_\infty$, and Nonlinear Approaches}, chapter~9.
\newblock Wiley, 2006.
\newblock \doi{10.1002/0470045345}.

\bibitem[Sontag(1998)]{sontag1990}
Eduardo~D. Sontag.
\newblock \emph{Mathematical Control Theory: Deterministic Finite Dimensional Systems}.
\newblock Springer-Verlag, 1st edition, 1998.
\newblock \doi{978-1-4612-0577-7}.

\bibitem[Subasi and Demirekler(2014)]{SUBASI2014}
Yuksel Subasi and Mubeccel Demirekler.
\newblock Quantitative measure of observability for linear stochastic systems.
\newblock \emph{Automatica}, 50\penalty0 (6):\penalty0 1669--1674, 2014.
\newblock \doi{https://doi.org/10.1016/j.automatica.2014.04.008}.

\bibitem[Tenny and Rawlings(2002)]{tenny2002}
Matthew~J. Tenny and James~B. Rawlings.
\newblock Efficient moving horizon estimation and nonlinear model predictive control.
\newblock In \emph{Proc. of the 2002 American Control Conf.}, volume~6, pages 4475--4480, 2002.
\newblock \doi{10.1109/ACC.2002.1025355}.

\bibitem[Tichavsky et~al.(1998)Tichavsky, Muravchik, and Nehorai]{Tichavsky1998}
Petr Tichavsky, Carlos~H. Muravchik, and Arye Nehorai.
\newblock Posterior {C}ramér-{R}ao bounds for discrete-time nonlinear filtering.
\newblock \emph{IEEE Trans. on Signal Processing}, 46\penalty0 (5):\penalty0 1386--1396, 1998.
\newblock \doi{10.1109/78.668800}.

\end{thebibliography}

\end{document}